\RequirePackage{fix-cm}
\documentclass[english,smallextended]{svjour3}       
\smartqed  

\usepackage{graphicx}
\usepackage{amssymb}
\usepackage{amsmath}
\usepackage{color}
\usepackage{soul}
\begin{document}

\title{Controlled-NOT gate sequences for mixed spin qubit architectures in a noisy environment
}

\titlerunning{CNOT gate sequences for mixed spin qubit architectures in a noisy environment}        

\author{E. Ferraro\and M. Fanciulli\and M. De Michielis}


\institute{E. Ferraro 
              \and 
              M. De Michielis              
             \at CNR-IMM, Agrate Brianza Unit, Via Olivetti 2, I-20864 Agrate Brianza (MB), Italy \\
             \email{elena.ferraro@mdm.imm.cnr.it}           
               \and  
              M. Fanciulli \at 
              CNR-IMM, Agrate Brianza Unit, Via Olivetti 2, I-20864 Agrate Brianza (MB), Italy\at 
              Dipartimento di Scienza dei Materiali, University of Milano Bicocca, Via R. Cozzi, 55, I-20125 Milano, Italy\\
              }           

\date{Received: date / Accepted: date}

\maketitle

\begin{abstract}
Explicit controlled-NOT gate sequences between two qubits of different types are presented in view of applications for large-scale quantum computation. Here, the building blocks for such composite systems are qubits based on the electrostatically confined electronic spin in semiconductor quantum dots. For each system the effective Hamiltonian models expressed by only exchange interactions between pair of electrons are exploited in two different geometrical configurations. A numerical genetic algorithm that takes into account the realistic physical parameters involved is adopted. Gate operations are addressed by modulating the tunneling barriers and the energy offsets between different couple of quantum dots. Gate infidelities are calculated considering limitations due to unideal control of gate sequence pulses, hyperfine interaction and charge noise. 
\keywords{Quantum computation architectures and implementations\and Quantum dots \and Noise \and Quantum gate sequences}
 \PACS{03.67.Lx, 73.21.La, 03.67.Ac}
\end{abstract}

\section{Introduction}
Semiconductor quantum dot (QD) systems are considered an attractive and scalable platform for quantum computing. The choice of the most suitable qubit and the consequent implementation of universal one and two-qubit gates are the main challenges for the realization of efficient quantum computers. A spin based quantum computer can be created exploiting different architectures that have been realized with great reliability in III-V compounds such as GaAs \cite{Medford-2013,Koppens-2006,Petta-2005} as well as in Si \cite{Veldhorst-2014,Kawakami-2014,Maune-2012}, besides theoretically investigated. A significant improvement in coherence times comes from the realization of QDs in Si, which can be isotopically purified.

The single localized electron spin in a quantum dot \cite{Morton-2011} is the simple and natural candidate to reproduce a qubit thanks to the direct identification that it is possible to envisage between the logical states $\{|0\rangle,|1\rangle\}$ and the angular momentum states of a particle with spin $1/2$. However also more advanced and most promising architectures have been proposed. In the double QD singlet-triplet qubit \cite{Levy-2002} the logical states are encoded with the singlet and triplet states of two electrons spatially separated in two QDs. The double QD hybrid qubit \cite{Shi-2012} combines spin and charge and the qubit is realized with three electrons arranged in two QDs. For all of these architectures, protocols for the implementation of single and two-qubit gates have been proposed \cite{Loss-1998,DiVincenzo-2000,Mehl-2014,Doherty-2013}. Also experiments have been realized in some cases ranging from the realization of single qubit gates to the two-qubit gates for the single spin qubit \cite{Veldhorst-2015}. 

The aim of the present paper is to merge all the informations coming from the architectures discussed until now in order to propose alternative sequence schemes for the realization of two-qubit controlled-NOT (CNOT) gates in mixed spin qubit architectures for large-scale quantum computation, following an analogous approach presented in a recent published paper \cite{Mehl-2015}. The increasing interest in this topic is also witnessed by experimental work \cite{Nakajima-2016}. In particular, the attention is focused on the interaction of the double QDs hybrid qubit, deeply experimentally tested \cite{Koh-2012,Kim-2012,Kim-2015}, with the single spin qubit in one case and with the singlet-triplet qubit in the other. It was demonstrated that the hybrid qubit up to now represents the most valuable candidate for several reasons. First of all due to its advantages in terms of the field control, the only exchange interaction mechanism among electrons removes the need for an inhomogeneous field allowing an all-electrical control of the system via gate electrodes. Moreover it assures an higher protection from hyperfine interactions of the singlet and triplet state in one of the two dots and guarantees a more compact fabrication, requiring the fabrication of two dots instead of three, as the qubit proposed by DiVincenzo \cite{DiVincenzo-2000} requires, where three electrons are arranged in three dots. All of the advantages of the hybrid qubit are combined with the convenience in using the single spin and the singlet-triplet qubit architectures, that although need a control via magnetic field or electron spin resonance (ESR) techniques, assure long coherence times with respect to gate operations. 

The paper is organized as follows. Section 2 presents the different spin qubit architectures, relying in particular on the effective Hamiltonian models and the definition of the logical states. In section 3 the CNOT gate sequences for the different geometrical configurations are derived adopting a search genetic algorithm, while in section 4 a gate fidelities analysis about the limitations of the results presented due to deviations from ideal pulses, hyperfine interaction and charge noise is discussed. Finally some concluding remarks are summarized.  

\section{Spin qubit architectures}
In this section the qubit architectures relying on the spin of electrons in electrostatically defined QDs are described. All the Hamiltonian models are presented in $\hbar$ units.

\subsection{Quantum dot single-spin qubit}
The QD single-spin qubit is realized using the spin of a single electron confined in a QD. The logical qubit basis is simply defined by the two spin eigenstates $|0\rangle\equiv|\!\uparrow\rangle$ and $|1\rangle\equiv|\!\downarrow\rangle$, where $|\!\uparrow\rangle(|\!\downarrow\rangle)$ correspond to the angular momentum state with $S=\frac{1}{2}$, $S_z=\frac{1}{2}(-\frac{1}{2})$. Universal qubit control of the spin qubit is achieved adopting magnetic fields pulses. From a practical point of view the approaches implemented are based on: electron spin resonance (ESR) techniques using local AC magnetic fields \cite{Morton-2011}, the application of a global magnetic field that allow local manipulation or through all-electrical manipulation via AC electric fields in a magnetic field gradient \cite{Kawakami-2014}. Although the manipulation requires sophisticated techniques involving magnetic fields or gradient of them, such spin qubit has a great advantage represented by long coherence times of the order of milliseconds \cite{Veldhorst-2014}. The Hamiltonian model is given by 
\begin{equation}
H=\frac{1}{2}E_z\sigma^z,
\end{equation}
where $\sigma_z$ is the Pauli operator and $E_z=g\mu_BB_z$ is the Zeeman energy associated to the magnetic field $B$ in the $z$ direction with $g$ the electron g-factor and $\mu_B$ the Bohr magneton.

\subsection{Double quantum dot singlet-triplet qubit}
The QD singlet-triplet qubit is created from two electrons in two QDs, ideally spatially separated. The logical states are defined by a superposition of two-particle spin singlet and triplet states, that are $|0\rangle\equiv|S\rangle$ and $|1\rangle\equiv|T_0\rangle$, where each QD is occupied with one electron. It is an external magnetic field that removes the $|T_-\rangle$ and $|T_+\rangle$ branches. $|S\rangle$, $|T_0\rangle$ and $|T_{\pm}\rangle$ are respectively the singlet and triplet states of a pair of electrons given by
\begin{align}\label{st1}
&|S\rangle=\frac{1}{\sqrt{2}}(|\!\uparrow\downarrow\rangle-|\!\downarrow\uparrow\rangle), \quad |T_0\rangle=\frac{1}{\sqrt{2}}(|\!\uparrow\downarrow\rangle+|\!\downarrow\uparrow\rangle),\nonumber\\
&|T_-\rangle=|\!\downarrow\downarrow\rangle, \quad |T_+\rangle=|\!\uparrow\uparrow\rangle.
\end{align} 
Spin rotations have been performed acting on the exchange coupling $J_{12}$ between the two electrons controlling via electrical potential the energy detuning between the two QDs. In addition a local magnetic field gradient is necessary to achieve arbitrary qubit rotations. The Hamiltonian model 
\begin{equation}
H=\frac{1}{2}\Delta E_z(\sigma_{1}^z-\sigma_{2}^z)+\frac{1}{4}J_{12}\boldsymbol{\sigma}_{1}\cdot\boldsymbol{\sigma}_{2}
\end{equation}
contains the exchange interaction between the two electrons, described by the Pauli matrices $\boldsymbol{\sigma}_1$ and $\boldsymbol{\sigma}_2$, through the coupling constant $J_{12}$, beyond the Zeeman term where a magnetic field gradient $\Delta E_z=1/2(E_{1}^z-E_{2}^z)$ is introduced between the two QDs. This architecture allows fast readout and manipulation, however experimentally the challenge is represented by the creation of the local magnetic gradient \cite{XianWu-2014,Barnes-2016,Pioro-2008,Maune-2012,Gonzalez-2012}. A valid strategy to overcome such task is represented by the use of a micro-magnet in close proximity \cite{XianWu-2014,Pioro-2008}.

\subsection{Double quantum dot hybrid qubit}
The double QD hybrid qubit owes its name to the fact that is an \emph{hybrid} of spin and charge \cite{Shi-2012}. It is composed by two QDs in which three electrons have been confined with all-electrical control via gate electrodes. The logical states coded using the $S=\frac{1}{2}$ and $S_z=\frac{1}{2}$ subspace of three electrons, have been defined by adopting combined singlet and triplet states of a pair of electrons occupying one dot with the states of the single electron occupying the other. The logical states have been expressed by $|0\rangle\equiv|S\rangle|\!\uparrow\rangle$ and $|1\rangle\equiv\sqrt{\frac{1}{3}}|T_0\rangle|\!\uparrow\rangle-\sqrt{\frac{2}{3}}|T_+\rangle|\!\downarrow\rangle$ where $|S\rangle$, $|T_0\rangle$ and $|T_{\pm}\rangle$ are defined in Eq.(\ref{st1}). The effective Hamiltonian model involving only exchange interaction terms among couple of electrons for a single and two qubits was derived in Ref.\cite{Ferraro-2014} and in Ref.\cite{Ferraro-2015-qip}, respectively. The universal set of gates for this qubit architecture is presented in Ref.\cite{DeMichielis-2015} and an alternative communication strategy between hybrid qubits is given in Ref.\cite{Ferraro-2015-prb}. A complete and detailed design of a scalable architecture based on this qubit type is reported in Ref.\cite{Rotta-2016}.
For the single hybrid qubit the Hamiltonian is equal to 
\begin{equation}
H=\frac{1}{2}E_z(\sigma_{1}^z+\sigma_{2}^z+\sigma_{3}^z)+\frac{1}{4}J_{12}\boldsymbol{\sigma}_{1}\cdot\boldsymbol{\sigma}_{2}+\frac{1}{4}J_{13}\boldsymbol{\sigma}_{1}\cdot\boldsymbol{\sigma}_{3}+\frac{1}{4}J_{23}\boldsymbol{\sigma}_{2}\cdot\boldsymbol{\sigma}_{3}.
\end{equation} 
The key advantage of this architecture is that the control and manipulation of the qubit is all electrical, on the other hand the disadvantage is represented by shorter coherence times that are of the order of tens of nanoseconds \cite{Koh-2012,Kim-2012,Kim-2015}.

\section{CNOT gates in mixed architectures}
This section is devoted to the derivation of the CNOT gate sequences in mixed systems in which the three coded qubits presented in the previous section are interconnected in different geometrical configurations. All the sequences derived are based on the control via electrical manipulation of the exchange coupling constants, moreover magnetic field gradients are required. For the first mixed architectures composed by the single spin with the hybrid qubit the magnetic field is modified dynamically, while for the singlet-triplet with the hybrid qubit mixed architecture the magnetic field gradient is kept constant.

\subsection{Quantum dot single-spin qubit and double quantum dot hybrid qubit}

\begin{figure}[h]
\begin{center}
\includegraphics[width=0.48\textwidth]{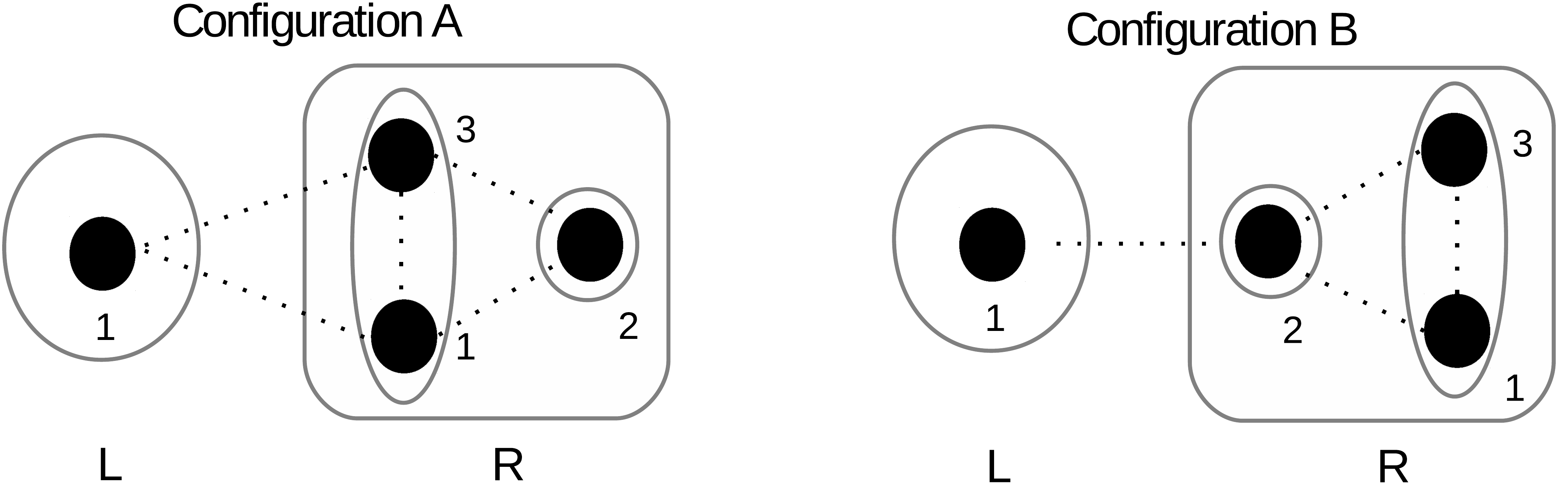}
\end{center}
\caption{Quantum dot single-spin qubit (L) and double quantum dot hybrid qubit (R): configurations A and B.}\label{single_hybrid}
\end{figure}

The mixed architecture composed by the QD single-spin qubit $L$ (left) interacting with a double-QD hybrid qubit $R$ (right) depicted in Fig. \ref{single_hybrid} is described by an effective Hamiltonian model (see Ref.\cite{Ferraro-2014,Ferraro-2015-qip}) where the free Hamiltonians $H_L$ and $H_R$ are added to an interaction term $H_{LR}^A(H_{LR}^B)$ where the superscript A(B) refers to the geometrical configuration under investigation:
\begin{align}\label{Hsingle}
&H=H_L+H_R+H_{LR}^{A,B}\nonumber\\
&H_L=\frac{1}{2}E_z\sigma_{1_L}^z\nonumber\\
&H_R=\frac{1}{4}J_{1_R2_R}\boldsymbol{\sigma}_{1_R}\cdot\boldsymbol{\sigma}_{2_R}+\frac{1}{4}J_{1_R3_R}\boldsymbol{\sigma}_{1_R}\cdot\boldsymbol{\sigma}_{3_R}+\nonumber\\
&+\frac{1}{4}J_{2_R3_R}\boldsymbol{\sigma}_{2_R}\cdot\boldsymbol{\sigma}_{3_R}+\frac{1}{2}E_z(\sigma_{1_R}^z+\sigma_{2_R}^z+\sigma_{3_R}^z)\nonumber\\
&H_{LR}^A=\frac{1}{4}J_{1_L1_R}\boldsymbol{\sigma}_{1_L}\cdot\boldsymbol{\sigma}_{1_R}+\frac{1}{4}J_{1_L3_R}\boldsymbol{\sigma}_{1_L}\cdot\boldsymbol{\sigma}_{3_R}\nonumber\\
&H_{LR}^B=\frac{1}{4}J_{1_L2_R}\boldsymbol{\sigma}_{1_L}\cdot\boldsymbol{\sigma}_{2_R}.
\end{align}
Since the two QDs composing the hybrid qubit are not symmetric two configurations are considered. The configuration A corresponds to the case in which the QD single-spin qubit is put into direct connection with the QD of the hybrid qubit containing two electrons. Analogously the configuration B takes into account the direct interaction of the QD single-spin qubit with the QD of the hybrid qubit containing one electron. The global magnetic field $E_z$ acts on the whole system, the coupling constants $J_{1_R2_R}, J_{1_R3_R}, J_{2_R3_R}$ represent the exchange interaction among the electronic spins in the hybrid qubit $R$, while $J_{1_L1_R}, J_{1_L3_R}, J_{1_L2_R}$ are the inter-qubit exchange couplings among electrons belonging to qubit $L$ and $R$. The explicit expressions for all the coupling constants are presented in Appendix A.

Before considering the Hilbert space of the whole system, we introduce the logical basis state of the QD single-spin qubit that is:
\begin{align}\label{single}
&|0_L\rangle\equiv|\!\uparrow\rangle\nonumber\\
&|1_L\rangle\equiv|\!\downarrow\rangle.
\end{align}
Moreover the basis state of the Hilbert space of the double-QD hybrid qubit, derived adopting the quantum angular momentum theory on a system of three spin $\frac{1}{2}$ particles, is:
\begin{align}\label{double}
&|0_R\rangle\equiv\frac{1}{\sqrt{2}}(|\!\uparrow\uparrow\downarrow\rangle-|\!\downarrow\uparrow\uparrow\rangle)\nonumber\\
&|1_R\rangle\equiv\frac{1}{\sqrt{6}}(|\!\uparrow\uparrow\downarrow\rangle+|\!\downarrow\uparrow\uparrow\rangle)-\sqrt{\frac{2}{3}}|\!\uparrow\downarrow\uparrow\rangle\nonumber\\
&|u_{-\frac{1}{2}}\rangle=\frac{1}{\sqrt{2}}(|\!\uparrow\downarrow\downarrow\rangle-|\!\downarrow\downarrow\uparrow\rangle)\nonumber\\
&|v_{-\frac{1}{2}}\rangle=\frac{1}{\sqrt{6}}(|\!\uparrow\downarrow\downarrow\rangle+|\!\downarrow\downarrow\uparrow\rangle)-\sqrt{\frac{2}{3}}|\!\downarrow\uparrow\downarrow\rangle\nonumber\\
&|Q_{\frac{3}{2}}\rangle=|\!\uparrow\uparrow\uparrow\rangle\nonumber\\
&|Q_{\frac{1}{2}}\rangle=\frac{1}{\sqrt{3}}(|\!\uparrow\uparrow\downarrow\rangle+|\!\uparrow\downarrow\uparrow\rangle+|\!\downarrow\uparrow\uparrow\rangle)\nonumber\\
&|Q_{-\frac{1}{2}}\rangle=\frac{1}{\sqrt{3}}(|\!\downarrow\downarrow\uparrow\rangle+|\!\downarrow\uparrow\downarrow\rangle+|\!\uparrow\downarrow\downarrow\rangle)\nonumber\\
&|Q_{-\frac{3}{2}}\rangle=|\!\downarrow\downarrow\downarrow\rangle,
\end{align}
where $|0_R\rangle$ and $|1_R\rangle$ are the chosen logical states for the qubit $R$, $|u_{-\frac{1}{2}}\rangle$ and $|v_{-\frac{1}{2}}\rangle$ span the $S=\frac{1}{2}$, $S_z=-\frac{1}{2}$ spin subspace and $|Q_i\rangle$ with $i=-\frac{3}{2},-\frac{1}{2},\frac{1}{2},\frac{3}{2}$ are the $S=\frac{3}{2}$ quadruplet states.

The composite system of the two qubits, that is four electronic spins, is described by a global Hilbert space $\mathcal{H}$ of dimension $2^4$. The computational space exploited for the CNOT sequences search is a subspace of $\mathcal{H}$ and it is spanned by an eight-dimensional basis in the subspaces with total angular momentum equal to $S=0$, $S_z=0$ and $S=1$, $S_z=0,1$. The basis states obtained composing the one qubit states (\ref{single}) and (\ref{double}) with appropriate Clebsch-Gordan coefficients are given by:
\begin{align}\label{basis1}
&|b_1\rangle=|0_L\rangle|0_R\rangle\nonumber\\
&|b_2\rangle=|0_L\rangle|1_R\rangle\nonumber\\
&|b_3\rangle=|1_L\rangle|0_R\rangle\nonumber\\
&|b_4\rangle=|1_L\rangle|1_R\rangle\nonumber\\
&|b_5\rangle=|0_L\rangle|u_{-\frac{1}{2}}\rangle\nonumber\\
&|b_6\rangle=|0_L\rangle|v_{-\frac{1}{2}}\rangle\nonumber\\
&|b_7\rangle=\frac{1}{\sqrt{2}}(|0_L\rangle|Q_{-\frac{1}{2}}\rangle-|1_L\rangle|Q_{\frac{1}{2}}\rangle)\nonumber\\
&|b_8\rangle=\frac{1}{2}(\sqrt{3}|1_L\rangle|Q_{\frac{3}{2}}\rangle-|0_L\rangle|Q_{\frac{1}{2}}\rangle).
\end{align}
Basis vectors $|b_1\rangle-|b_4\rangle$ are valid encoded states in which the CNOT operator has the usual form
\begin{equation} \label{Eq:CNOT}
CNOT=\left(\begin{array}{cccc}
1 & 0 & 0 & 0\\
0 & 1 & 0 & 0\\
0 & 0 & 0 & 1\\
0 & 0 & 1 & 0
\end{array}\right).
\end{equation}
Considering that $E_z,J_{1_R2_R},J_{1_R3_R},J_{2_R3_R}\gg J_{1_L1_R},J_{1_L3_R},J_{1_L2_R}$ and that the global magnetic field $E_z$ and the exchange interactions of the hybrid qubit cause single qubit time evolutions that will be neglected, the basis adopted for our calculation is limited to $\{|b_1\rangle-|b_6\rangle\}$. In order to obtain the exchange interaction sequences for the CNOT gates we developed a genetic search algorithm similar to the one described in Ref.\cite{Fong-2011}, which is a combination of a simplex-based and a genetic algorithms. The magnitudes of the exchange and magnetic fields pulses were chosen in advance, but their ordering and duration were determined by the genetic algorithms. At each iteration of the search algorithm sequences become closer to the global minimum, featuring a reduced number of exchange steps and minimum interaction time per step. More in detail the genetic approach that we have implemented iterates random changes on some population of gate configurations, each given by a list of interactions between adjacent electrons. At each iteration (generation) we augment the population with mutations and mating, and then apply natural selection that favors those configurations with low value of the objective function. The objective function is augmented by adding a gate penalty term which is simply the length of the sequence times a positive constant, the gate penalty parameter. Time-optimal sequences \cite{Haidong_PRA2005,Navin_PRA2001} are provided with this approach where time-optimality is assured by searching for sequences with few steps with shortest step times.
The objective function \cite{Fong-2011} is defined as:
\begin{equation}\label{func1}
f_{CNOT}=\sqrt{1-\frac{1}{4}\left|U_{(1,1)}+U_{(2,2)}+U_{(3,4)}+U_{(4,3)}\right|}.
\end{equation}
The unitary evolution is given by $U(t)=e^{-iHt}$ where $H$ is the Hamiltonian of the whole system given in Eq.(\ref{Hsingle}) and $U_{(i,j)}$ appearing in Eq.(\ref{func1}) are the matrix elements of the CNOT gate in the encoded states in the $6\times 6$ subspace. The objective function is exactly equal to zero, that means that the gate sequence is optimal, when all the $U$ entering into Eq.(\ref{func1}) have modulus $1$ and a common phase. 

In the realistic situation investigated the unavoidable intra-dot interaction in the double QD hybrid qubit $J_{1_R3_R}$ is fixed and not tunable from the external, in fact it does not depend on the tunneling rates \cite{Ferraro-2015-qip}. We assumed $\max(J_{1_R2_R})=\max(J_{2_R3_R})=J^{max}$ and we set a realistic constant value for $J_{1_R3_R}$ = $J^{max}/2$ to model the ineffective control. Moreover, the exchange interactions $J_{1_R2_R}(t)$ and $J_{2_R3_R}(t)$ are assumed to have instantaneous turn-on and turn-off. 
The global magnetic field $E_z$ in real experiments can reach values of tens of $\mu$eV for external magnetic fields above hundreds of mT \cite{Yoneda-2014}. Also the value for $E_z$ is defined in units of $J^{max}$, that has been estimated for the cases under investigation to be 1 $\mu$eV (value compatible with that reported in Ref.\cite{DeMichielis-2015}), giving as a result $E_z=10J^{max}$. An additional magnetic field $E_z^{\ast}$ acting on the single spin qubit is included as an external parameter of control with $\max(E_z^{\ast})=J^{max}$. The variation of this additional magnetic field is done exploiting magnetic field gradients by moving on the electron between regions with different magnetic fields. Magnetic field gradients of 1.5 mT/nm are expected when QDs are embedded in an external magnetic field and located closer to an integrated micromagnet as shown in Ref \cite{Neumann-2015}. 
In these hypothesis we derive for both configurations A and B the sequences of unitary operations generated by exchange interactions, whose product gives the CNOT operation in the total angular momentum basis.

\begin{figure}[htp]
\centering
\hfill
\begin{minipage}[b]{.4\columnwidth}
  \centering
  \begin{tabular}{c c c c}
Step & Int & t[$h/J^{max}$] & t$[ns]$ \\
\hline
1 & $J_{1_R2_R}$ & 0.5928 & 2.4518 \\ 
2 & $J_{2_R3_R}$ & 0.5768 & 2.3856 \\
3 &  $J_{1_L3_R}$ & 0.0006 & 0.0025 \\ 
4 & $J_{1_R2_R}$ & 0.0047 & 0.0194 \\
5 & $E_z^{\ast}$ & 0.2513 & 1.0393 \\
6 &  $J_{1_L3_R}$ & 0.0004 & 0.0017 \\ 
7 & $J_{1_R2_R}$ & 0.5966 & 2.4675 \\
8 & wait & 0.0142 & 0.0587\\  
9 & $J_{2_R3_R}$ & 0.0310 & 0.1282\\ 
10 & wait & 0.4009 & 1.6581\\
11 & $E_z^{\ast}$ & 0.0140 & 0.0579\\ 
12 & $J_{2_R3_R}$ & 0.0140 & 0.0579\\ 
13 & $E_z^{\ast}$ & 0.0081 & 0.0335\\ 
 14 & wait & 1.8139 & 7.5023\\
\hline
\end{tabular}
\end{minipage}
\hfill
\begin{minipage}{.4\columnwidth}
  \centering
  \includegraphics[width=1.3\textwidth]{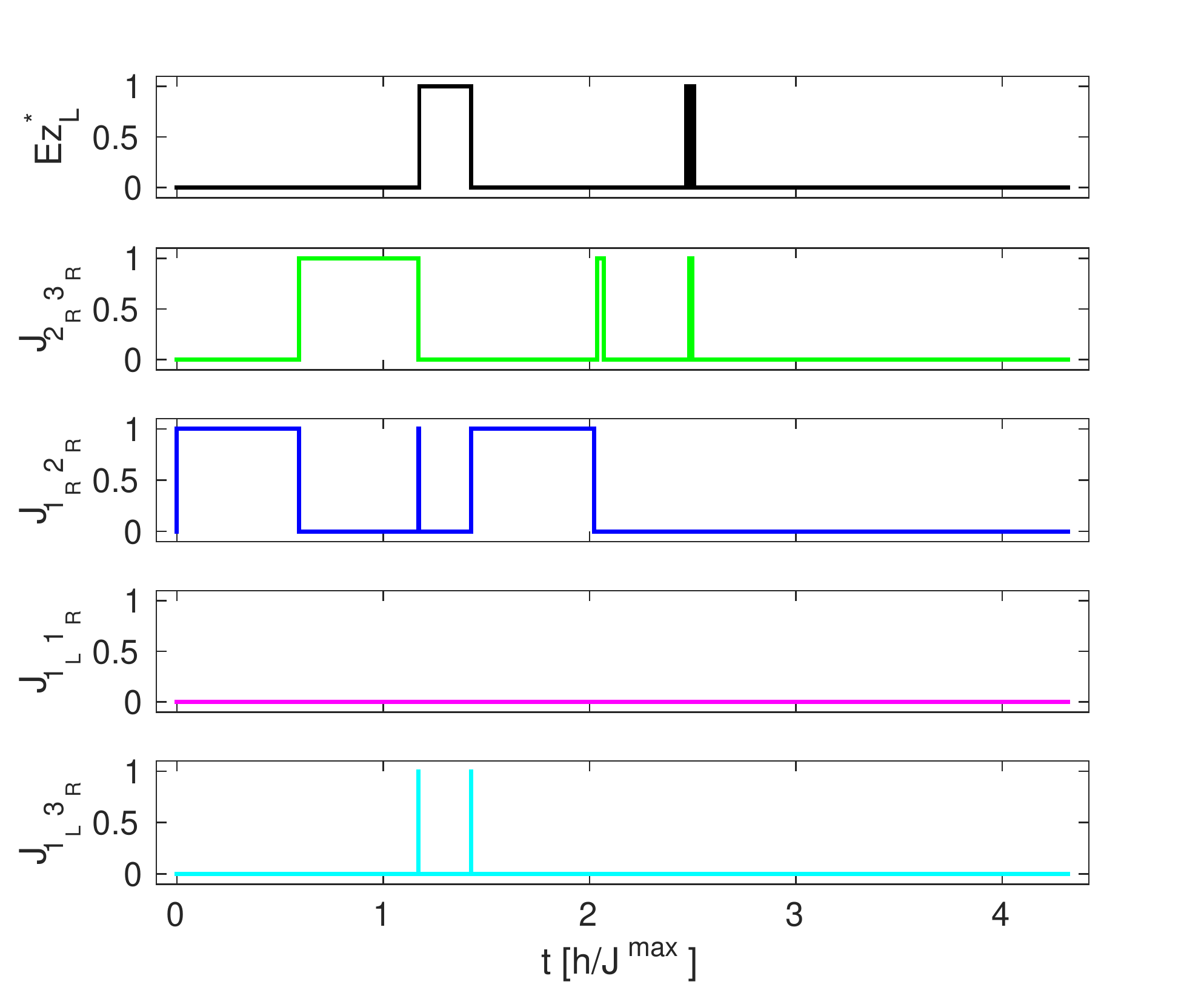}
\end{minipage}\hspace*{\fill}
\caption{Left: Gate sequence implementing a CNOT gate for the configuration A with fixed $J_{1_R3_R}$=$J^{max}/2$ and $E_z=10J^{max}$. The ``wait'' interaction represents only the fixed interactions $J_{1_R3_R}$ with no other interactions on. Times are in unit of $h/J^{max}$ in the third column and in physical units in correspondance to $J^{max}=$1 $\mu$eV in the fourth column. Right: Graphical representation}\label{Tab:SeqAndTimesCNOTv1A}
\end{figure}

The interaction sequence for a CNOT operation in the configuration A is calculated by using the search algorithm in the case of fixed intra-dot interaction for the double QD hybrid qubit. The resulting sequence is reported in Fig.\ref{Tab:SeqAndTimesCNOTv1A}. 

\begin{figure}[htp]
\centering
\hfill
\begin{minipage}[b]{.4\columnwidth}
  \centering
  \begin{tabular}{c c c c}
Step & Int & t[$h/J^{max}$] & t$[ns]$ \\
\hline
1 & $J_{1_R2_R}$ & 0.5526 & 2.2856 \\  
2 & wait & 0.0093 & 0.0385 \\
3 & $E_z^{\ast}$ & 0.4935 & 2.0411 \\
4 & $J_{1_R2_R}$ & 0.0014 & 0.0058 \\
5 & wait & 0.0235 & 0.0972 \\ 
6 & $J_{1_L2_R}$ & 0.0266 & 0.1100 \\
7 & wait & 0.0086 & 0.0356 \\ 
8 & $J_{2_R3_R}$ & 0.0068 & 0.0281\\
9 & wait & 1.4444 & 5.9740\\
10 & $J_{2_R3_R}$ & 0.0045 & 0.0186\\ 
11 & wait & 0.0217 & 0.0898\\
12 & $J_{1_R2_R}$ & 0.0186 & 0.0769\\ 
13 & wait & 0.0404 & 0.1671\\
14 & $J_{1_R2_R}$ & 0.0097 & 0.0401\\
15 & $J_{1_L2_R}$ & 0.3834 & 1.586\\ 
16 & wait & 0.0452 & 0.1869\\
17 & $J_{1_L2_R}$ & 2.5885 & 10.706\\ 
\hline
\end{tabular}
\end{minipage}
\hfill
\begin{minipage}{.4\columnwidth}
  \centering
  \includegraphics[width=1.3\textwidth]{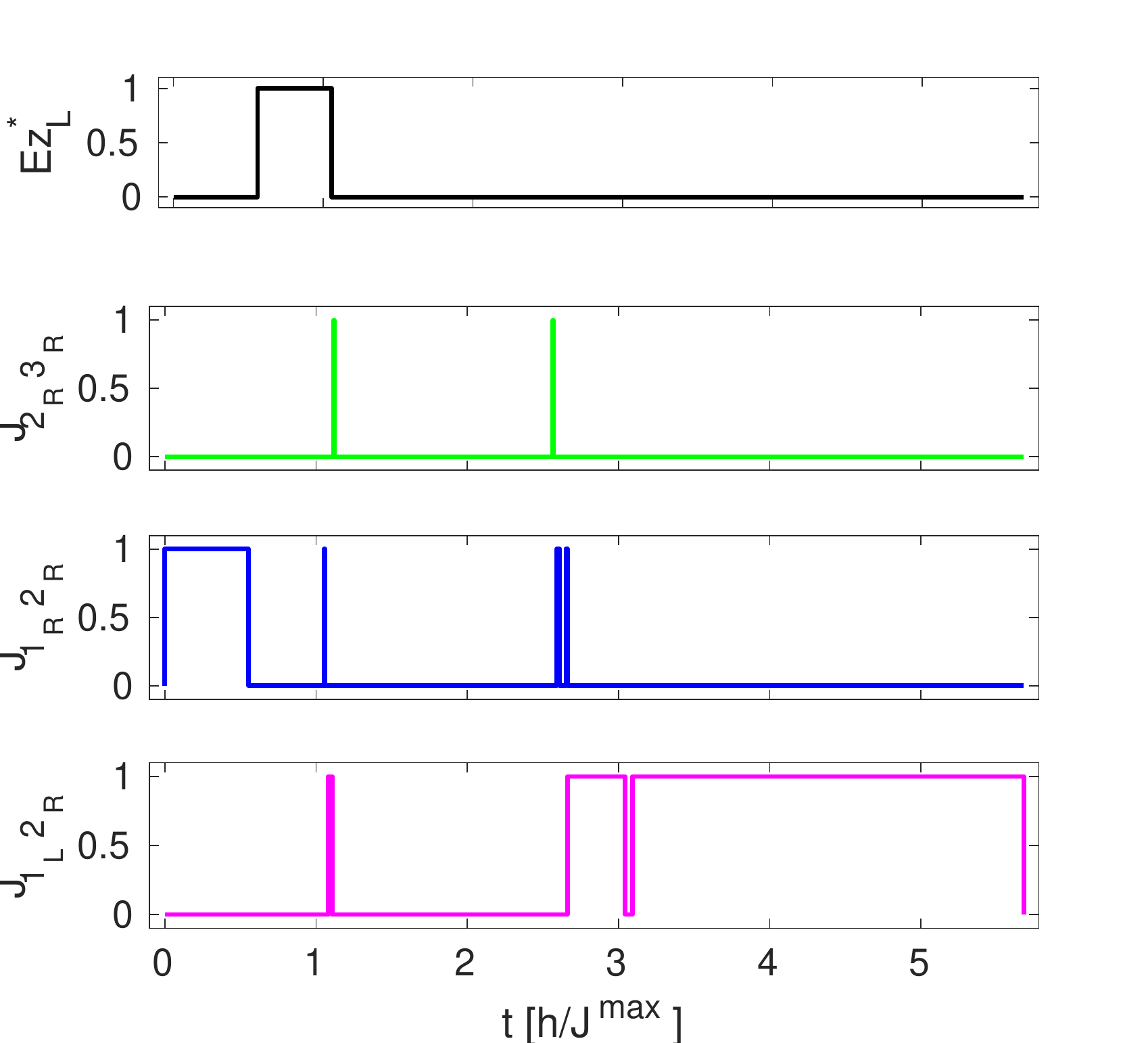}
\end{minipage}\hspace*{\fill}
\caption{As Fig.\ref{Tab:SeqAndTimesCNOTv1A} but for the configuration B.}\label{Tab:SeqAndTimesCNOTv1B}
\end{figure}

Analogously the interaction sequence for the configuration B is presented in Fig.\ref{Tab:SeqAndTimesCNOTv1B}.

\subsection{Double quantum dot singlet-triplet qubit and double quantum dot hybrid qubit}

\begin{figure}[h]
\begin{center}
\includegraphics[width=0.48\textwidth]{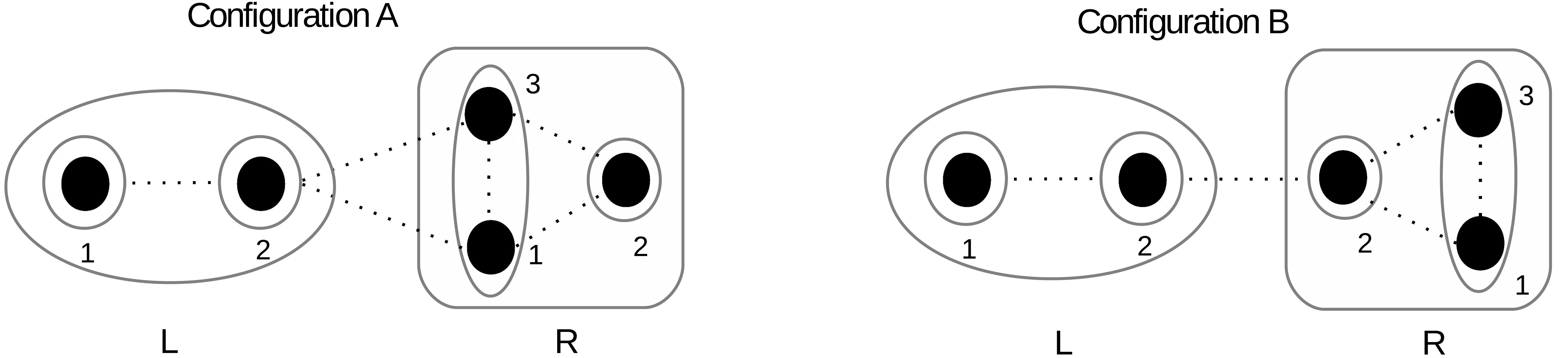}
\end{center}
\caption{Double quantum dot singlet-triplet qubit (L) and double quantum dot hybrid qubit (R): configurations A and B.}\label{st_hybrid}
\end{figure}

The Hamiltonian model of the mixed system composed by a double QD singlet-triplet qubit $L$ with a double-QD hybrid qubit $R$ is given by the sum of the free Hamiltonians $H_L$ and $H_R$ with the interaction term:
\begin{align}\label{Hst}
&H=H_L+H_R+H_{LR}^{A,B}\nonumber\\
&H_L=\frac{1}{2}(E_z+\tilde{E}_z)(\sigma_{1_L}^z+\sigma_{2_L}^z)+\frac{1}{4}J_{1_L2_L}\boldsymbol{\sigma}_{1_L}\cdot\boldsymbol{\sigma}_{2_L}\nonumber\\
&H_R=\frac{1}{4}J_{1_R2_R}\boldsymbol{\sigma}_{1_R}\cdot\boldsymbol{\sigma}_{2_R}+\frac{1}{4}J_{1_R3_R}\boldsymbol{\sigma}_{1_R}\cdot\boldsymbol{\sigma}_{3_R}+\nonumber\\
&+\frac{1}{4}J_{2_R3_R}\boldsymbol{\sigma}_{2_R}\cdot\boldsymbol{\sigma}_{3_R}+\frac{1}{2}E_z(\sigma_{1_R}^z+\sigma_{2_R}^z+\sigma_{3_R}^z)\nonumber\\
&H_{LR}^A=\frac{1}{4}J_{2_L1_R}\boldsymbol{\sigma}_{2_L}\cdot\boldsymbol{\sigma}_{1_R}+\frac{1}{4}J_{2_L3_R}\boldsymbol{\sigma}_{2_L}\cdot\boldsymbol{\sigma}_{3_R}\nonumber\\
&H_{LR}^B=\frac{1}{4}J_{2_L2_R}\boldsymbol{\sigma}_{2_L}\cdot\boldsymbol{\sigma}_{2_R}.
\end{align}
Once again, as in the previous case, due to the asymmetries in the QDs composing the hybrid qubit, two different configurations shown in Fig. \ref{st_hybrid} are considered that correspond to the two interaction Hamiltonian models $H_{LR}^A$ and $H_{LR}^B$. A global magnetic field $E_z$ acts on the whole system, in addition a contribution to the Zeeman energy comes from $\tilde{E}^z$, that represents the magnetic field variations at the double QD singlet-triplet qubit. The explicit expressions for all the coupling constants are presented in Appendix A.

The logical states $|0_L\rangle$ and $|1_L\rangle$ of the singlet-triplet qubit are defined as:
\begin{align}\label{singlet2}
&|0_L\rangle\equiv|S\rangle\nonumber\\
&|1_L\rangle\equiv|T_0\rangle,
\end{align}
where $|S\rangle$ and $|T_0\rangle$ are respectively the singlet and triplet states of two spins $\frac{1}{2}$, given in Eq.(\ref{st1}), while the logical states of the double QD hybrid qubit are given in Eqs.(\ref{double}).

The composite system of the two qubits, that is five electronic spins, is described by a nine-dimensional basis in the subspaces with total angular momentum equal to $S=\frac{1}{2}$, $S_z=\frac{1}{2}$ and $S=\frac{3}{2}$, $S_z=\frac{1}{2}$ obtained composing the one qubit states (\ref{singlet2}) and (\ref{double}) with appropriate Clebsch-Gordan coefficients:
\begin{align}\label{basis2}
&|b_1\rangle=|0_L\rangle|0_R\rangle\nonumber\\
&|b_2\rangle=|0_L\rangle|1_R\rangle\nonumber\\
&|b_3\rangle=|1_L\rangle|0_R\rangle\nonumber\\
&|b_4\rangle=|1_L\rangle|1_R\rangle\nonumber\\
&|b_5\rangle=|T_+\rangle|u_{-\frac{1}{2}}\rangle\nonumber\\
&|b_6\rangle=|T_+\rangle|v_{-\frac{1}{2}}\rangle\nonumber\\
&|b_7\rangle=\frac{1}{\sqrt{2}}|T_-\rangle|Q_{\frac{3}{2}}\rangle-\frac{1}{\sqrt{3}}|T_0\rangle|Q_{\frac{1}{2}}\rangle+\frac{1}{\sqrt{6}}|T_+\rangle|Q_{-\frac{1}{2}}\rangle\nonumber\\
&|b_8\rangle=\sqrt{\frac{2}{5}}|T_-\rangle|Q_{\frac{3}{2}}\rangle+\frac{1}{\sqrt{15}}|T_0\rangle|Q_{\frac{1}{2}}\rangle-\sqrt{\frac{8}{15}}|T_+\rangle|Q_{-\frac{1}{2}}\rangle\nonumber\\
&|b_9\rangle=|S\rangle|Q_{\frac{3}{2}}\rangle.
\end{align}
Using the same argument as before we note that $E_z,J_{1_L2_L},J_{1_R2_R},J_{1_R3_R},J_{2_R3_R}\gg \tilde{E}_z,J_{2_L1_R},J_{2_L3_R},J_{2_L2_R}$ and the qubit time evolution is restricted to the subspace $\{|b_1\rangle-|b_6\rangle\}$.

\begin{figure}[htp]
\centering
\hfill
\begin{minipage}[b]{\columnwidth}
  \centering
\begin{tabular}{c c c c|c c c c}
Step & Int & t[$h/J^{max}$] & t$[ns]$ & Step & Int & t[$h/J^{max}$] & t$[ns]$ \\
\hline
1 & $J_{2_L1_R}$  & 1.4076 & 5.8218 & 17 & wait & 0.0010 & 0.0041\\		 
2 & $J_{1_R2_R}$ & 1.7463 & 7.2227& 18 & $J_{1_L2_L}$ & 0.1416 & 0.5857\\ 
3 & $J_{2_L1_R}$ & 0.0357 & 0.1477& 19 & $J_{1_R2_R}$ & 0.0080 & 0.0331\\  
4 & wait  & 0.0272 & 0.1125& 20 & $J_{2_L3_R}$ & 0.7881 & 3.2596\\ 
5 & $J_{2_L3_R}$ & 0.0202 & 0.0835& 21 & $J_{1_R2_R}$ & 0.0297 & 0.1228\\ 
6 & $J_{1_R2_R}$ & 0.0259 & 0.1071& 22 & $J_{2_R3_R}$ & 0.0042 & 0.0174\\ 
7 & $J_{2_R3_R}$ & 0.0065 & 0.0269& 23 & wait & 0.8708 & 3.6016\\ 
8 & $J_{1_L2_L}$  & 1.4799 & 6.1209& 24 & $J_{2_L3_R}$ & 0.0028 & 0.0116\\ 
9 & $J_{2_L1_R}$ & 0.1691 & 0.6994& 25 & $J_{2_R3_R}$ & 1.3218 & 5.4670\\ 
10 & $J_{2_L3_R}$ & 0.0161 & 0.0666& 26 & $J_{1_R2_R}$ & 0.0268 & 0.1108\\ 
11 & $J_{1_L2_L}$ & 0.0107 & 0.0443& 27 & $J_{2_L3_R}$ & 0.2169 & 0.8971\\ 
12 & $J_{1_R2_R}$  & 0.4494 & 1.8587& 28 & $J_{2_R3_R}$ & 0.0379 & 0.1568\\ 
13 & wait & 2.9450 & 12.181& 29 & $J_{1_R2_R}$ & 1.1964 & 4.9483\\ 
14 & $J_{2_L3_R}$ & 0.0410 & 0.1696& 30 & $J_{2_L1_R}$ & 0.0602 & 0.2490\\ 
15 & $J_{1_L2_L}$ & 0.1423 & 0.5886& 31 & $J_{1_L2_L}$ & 0.2372 & 0.9811\\ 
16 & $J_{2_L3_R}$  & 0.2503 & 1.0352& 32 & $J_{1_R2_R}$ & 2.3079 & 9.5455\\ 
\hline
\end{tabular}
\end{minipage}
\hfill
\begin{minipage}{\columnwidth}
  \centering
  \includegraphics[width=0.5\textwidth]{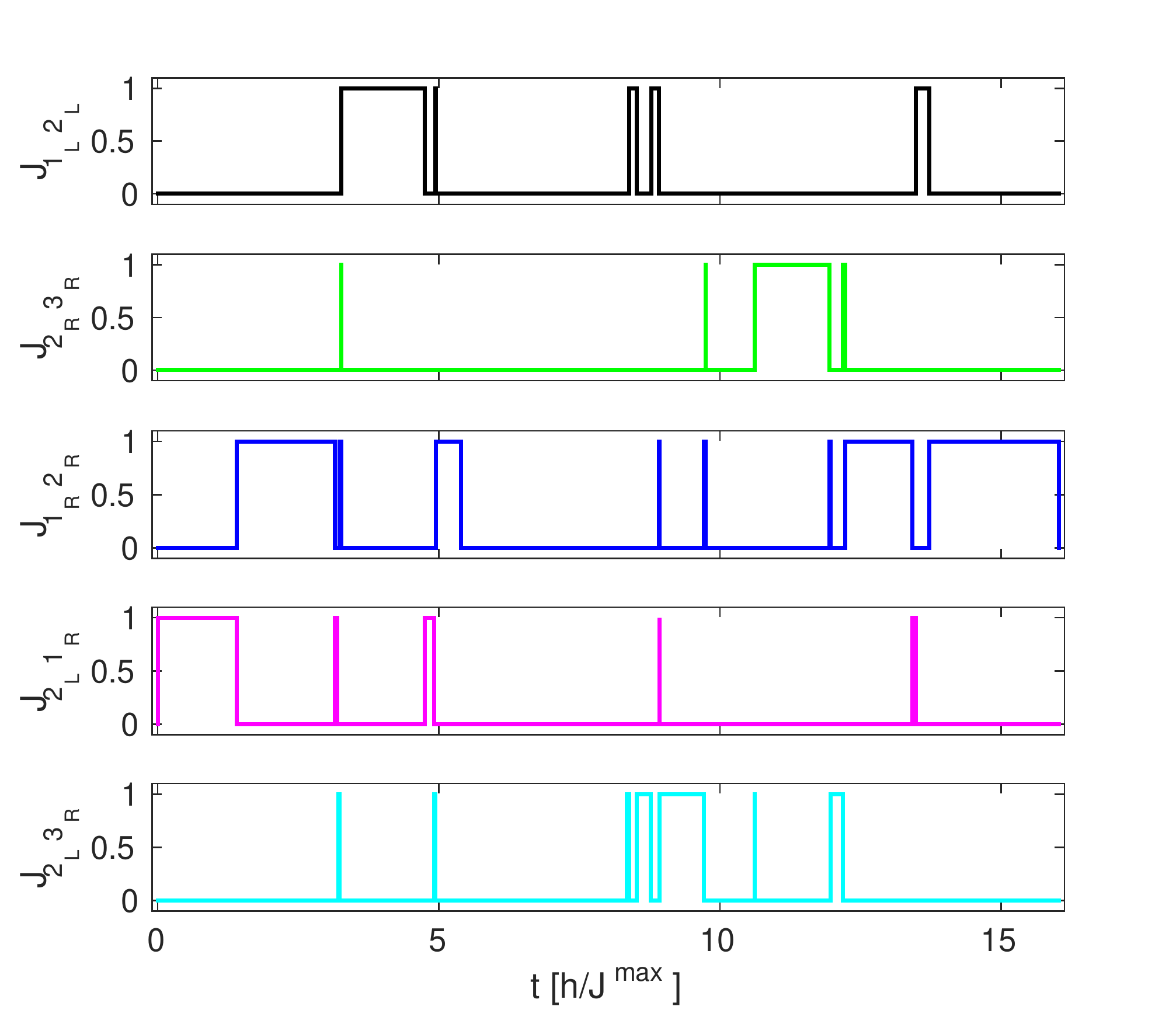}
\end{minipage}\hspace*{\fill}
\caption{Top: Gate sequence implementing a CNOT gate for the configuration A with fixed $J_{1_R3_R}$=$J^{max}/2$, $E_z=10J^{max}$ and $\tilde{E}_z=1J^{max}$. The ``wait'' interaction represents only the fixed interactions $J_{1_R3_R}$ with no other interactions on. Times are in unit of $h/J^{max}$ in the third column and in physical units in correspondance to $J^{max}=$1 $\mu$eV in the fourth column. Bottom: Graphical representation.}\label{Tab:SeqAndTimesCNOTv2A}
\end{figure}

\begin{figure}[htp]
\centering
\hfill
\begin{minipage}[b]{\columnwidth}
  \centering
\begin{tabular}{c c c c|c c c c}
Step & Int & t[$h/J^{max}$] & t$[ns]$ & Step & Int & t[$h/J^{max}$] & t$[ns]$ \\
\hline
1 & $J_{2_L2_R}$ & 0.7657 & 3.1669& 18 & wait & 1.3282 & 5.4934\\ 
2 & wait & 0.0329 & 0.1361& 19 & $J_{2_L2_R}$ & 0.2042 & 0.8446\\ 
3 & $J_{1_R2_R}$ & 0.7692 & 3.1814& 20 & $J_{1_L2_L}$ & 0.1896 & 0.7842\\ 
4 & $J_{2_R3_R}$ & 1.3622 & 5.6341& 21 &$J_{2_L2_R}$ & 0.6539 & 2.7045\\ 
5 & $J_{2_L2_R}$ & 0.7658 & 3.1674& 22 & wait & 0.3438 & 1.4220\\ 
6 & $J_{1_L2_L}$ & 0.3674 & 1.5196& 23 & $J_{2_L2_R}$ & 0.0024 & 0.0099\\ 
7 & $J_{2_L2_R}$ & 0.0154 & 0.0637& 24 & $J_{1_R2_R}$ & 0.3431 & 1.4191\\ 
8 & wait & 0.0073 & 0.0302& 25 & $J_{2_R3_R}$ & 2.9980 & 12.400\\ 
9 & $J_{1_R2_R}$ & 0.4400 & 1.8198& 26 & $J_{2_L2_R}$ & 0.1172 & 0.4847\\ 
10 & $J_{2_R3_R}$ & 0.1353 & 0.5596& 27 & $J_{1_R2_R}$ & 0.1714 & 0.7089\\ 
11 & $J_{1_L2_L}$ & 0.4220 & 1.7454& 28 & wait & 0.1257 & 0.5199\\ 
12 & $J_{2_L2_R}$ & 0.7984 & 3.3022& 29 & $J_{2_L2_R}$ & 0.0970 & 0.4012\\ 
13 & $J_{1_L2_L}$ & 1.0909 & 4.5120& 30 & $J_{2_R3_R}$ & 0.1531 & 0.6332\\ 
14 & $J_{2_L2_R}$ & 0.0003 & 0.0012& 31 & $J_{1_R2_R}$ & 0.7744 & 3.2029\\ 
15 & wait & 0.0903 & 0.3735& 32 & $J_{2_R3_R}$ & 0.7052 & 2.9167\\ 
16 & $J_{1_R2_R}$ & 0.4492 & 1.8579& 33 & wait & 0.1452 & 0.6006\\ 
17 & $J_{2_L2_R}$ & 0.1491 & 0.6167& 34 & $J_{1_L2_L}$ & 0.0270 & 0.1117\\ 
\hline
\end{tabular}
\end{minipage}
\hfill
\begin{minipage}{\columnwidth}
  \centering
  \includegraphics[width=0.5\textwidth]{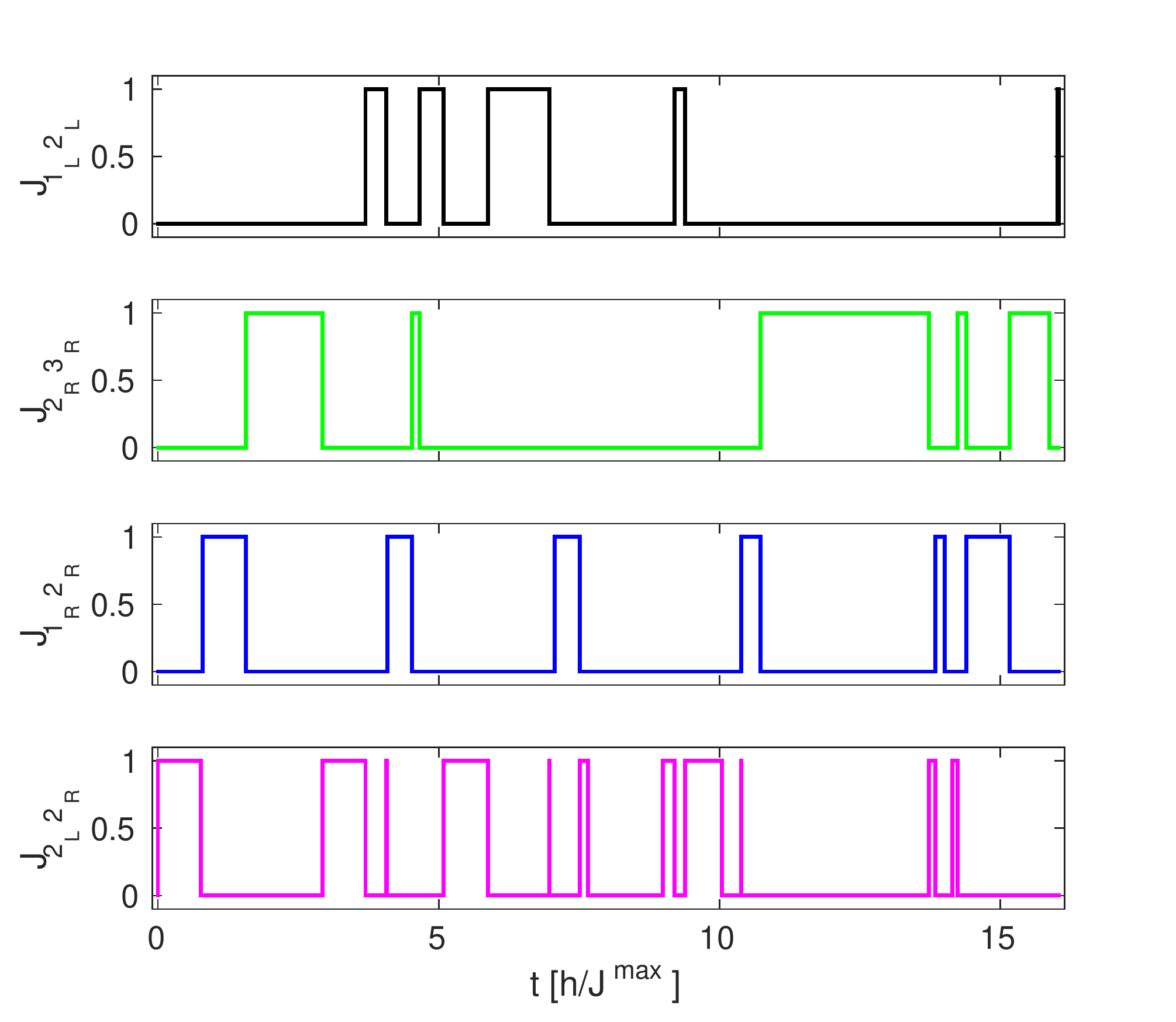}
\end{minipage}\hspace*{\fill}
\caption{As Fig.\ref{Tab:SeqAndTimesCNOTv2A} but for the configuration B.}\label{Tab:SeqAndTimesCNOTv2B}
\end{figure}

The CNOT gate sequence, obtained adopting the same objective function (\ref{func1}), for the configuration A (B) is presented in Fig.\ref{Tab:SeqAndTimesCNOTv2A} (\ref{Tab:SeqAndTimesCNOTv2B}). 

\section{Gate fidelity analysis for the CNOT sequences}
Non idealities must be included in the model to perform a good performance analysis in real systems. We account for noise sources such as time interval error (TIE) in gate sequences, hyperfine interaction with magnetic centers and charge noise. 
The figure of merit used to estimate the noise effects is the entanglement fidelity $F$ \cite{Nielsen-2000,Marinescu-2012}. 
A disturbed operation $U_{d}$ affects  
\begin{equation}
F= tr [\rho^{RS} \mathbf{1}_{R} \otimes (U_{i}^{-1}U_{d})_{S} \rho^{RS} \mathbf{1}_{R} \otimes (U_{d}^{-1}U_{i})_{S}]
\end{equation}
where $U_{i}$ is the ideal time evolution and $\rho^{RS}=|\psi\rangle\langle\psi|$ with $|\psi\rangle=1/2(|0000\rangle+|0110\rangle+|1001\rangle+|1111\rangle)$ represents a maximally entangled state in a double state space generated by two identical Hilbert spaces $R$ and $S$. 

We point out that all the CNOT sequences for the four different configurations are calculated with fidelity $F\geq 0.999998$ in absence of noise. This is an estimation of how much our CNOT sequences are far away from the exact CNOT in Eq.(\ref{Eq:CNOT}).


\subsection{Time interval error}
In our model, exchange coupling signal $J_{13}$ in the hybrid qubit is constant whereas others $J_{ij}$ are controlled in such a way that only one interaction is enabled at a time. As a result, taken two interaction signals, the fall edge of the first signal and the rise edge of the second one occur at the same time. Time Interval Error (TIE) of a signal edge is defined as the time deviation of that edge from its ideal position. The physical origins of this kind of error have to be searched in the non idealities and intrinsic delays of the pulse generator used to drive the qubits. Here, TIE is taken into account in the following way: the switch between the two interactions is still simultaneous but occurs at a different time with respect the undisturbed case. For each CNOT sequence step a TIE value is taken from a Gaussian distribution with zero mean and $\sigma$ standard deviation and added to the step time. After averaging over 10000 non-ideal sequences the resulting infidelity is calculated.  

\begin{figure}[h]
\begin{center}
\includegraphics[width=0.8\textwidth]{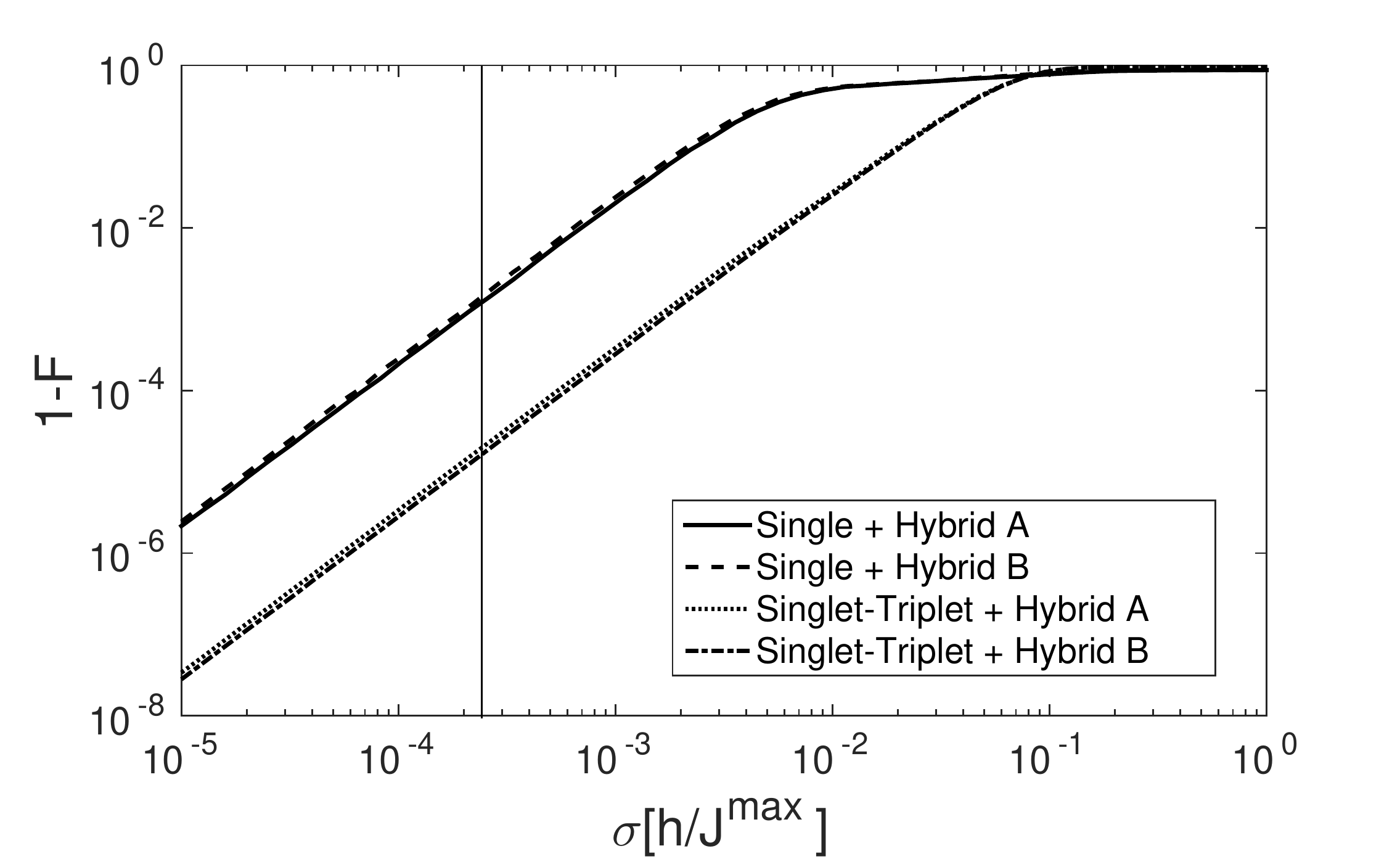}
\end{center}
\caption{\label{Fig:TIE} Gate infidelities for the CNOT gate sequences of Tabs. \ref{Tab:SeqAndTimesCNOTv1A}, \ref{Tab:SeqAndTimesCNOTv1B}, \ref{Tab:SeqAndTimesCNOTv2A}, \ref{Tab:SeqAndTimesCNOTv2B} due to TIE. Vertical line marks a typical experimental $\sigma$ of 1 ps, that correspond to $\sigma=0.25\times 10^{-3}h/J^{max}$ with $J^{max}$=1 $\mu$eV.}
\end{figure}

Figure \ref{Fig:TIE} shows the gate infidelities $1-F$ of the CNOT gate sequences reported in Tabs. \ref{Tab:SeqAndTimesCNOTv1A}, \ref{Tab:SeqAndTimesCNOTv1B}, \ref{Tab:SeqAndTimesCNOTv2A}, \ref{Tab:SeqAndTimesCNOTv2B} as a function of the $\sigma$ of TIE. 
All the systems have an infidelity that rises when the standard deviation of TIE increases. For very high values of $\sigma$, infidelity tends to saturate. Systems with singlet-triplet qubit are less sensitive to TIE with respect systems with single spin qubits whereas infidelities of systems with different hybrid qubit configuration do not differ too much.  

\subsection{Hyperfine interactions}
Hyperfine interactions generate magnetic field fluctuations at the QD sites. Due to their low frequency fluctuations can be treated as static during the CNOT gate sequences (Tabs. \ref{Tab:SeqAndTimesCNOTv1A}, \ref{Tab:SeqAndTimesCNOTv1B}, \ref{Tab:SeqAndTimesCNOTv2A}, \ref{Tab:SeqAndTimesCNOTv2B}) as done in Ref.\cite{Mehl-2014,Mehl-2015-2}. A random component $\delta E_{z}$ parallel to the external magnetic field is considered as the main error contribution with typical values for uncorrected nuclear spin baths of 100 neV for GaAs QDs \cite{Taylor-2007} and 3 neV in Si QDs \cite{Assali-2011}, that correspond to magnetic fields of 5 mT and 25 $\mu$T, respectively. These values are used as the rms of a Gaussian distribution for $\delta E_{z}$ at each QD, averaging over 10000 nuclear distributions with random $\delta E_{z}$. 
\begin{figure}[h]
\begin{center}
\includegraphics[width=0.8\textwidth]{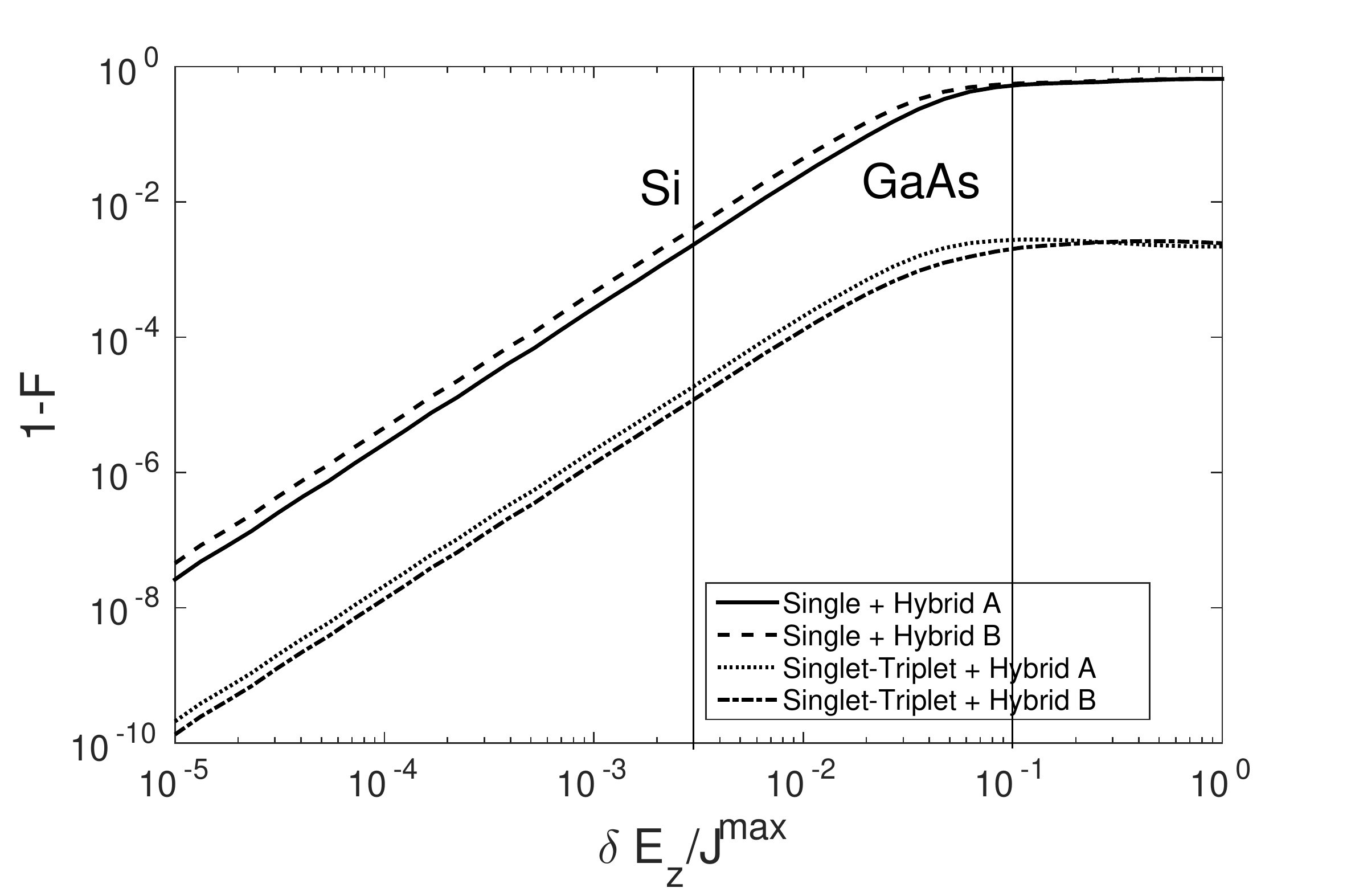}
\end{center}
\caption{\label{Fig:dEz} Gate infidelities for the CNOT gate sequences due to the effect of hyperfine interaction for the four different systems. Vertical lines mark typical $\delta E_{z}$ for Si and GaAs QDs when $J^{max}$=1 $\mu$eV.}
\end{figure}
Figure \ref{Fig:dEz} shows the gate infidelities $1-F$ of the CNOT gate sequences as a function of the $\delta E_{z}$ disturb for the different systems considered. Systems have infidelity that increases with respect to the intensity of the random component $\delta E_{z}$, saturating at high values of $\delta E_{z}$. As in Fig. \ref{Fig:TIE}, systems with singlet-triplet qubit are less sensitive to $\delta E_{z}$ with respect to systems with single spin qubits whereas infidelities of systems with different hybrid qubit configuration slightly differ. Obviously, when intrinsic Si is considered, isotopically purified $^{28}$Si is the ultimate host material to cancel the effects of hyperfine coupling. 

\subsection{Charge noise}
Defects in the device may trap and emit charges. These are stochastic processes leading to 1/f noise in large devices, where several defects are active, and to random telegraph noise (RTN) in scaled devices where a single trap is active. RTN generates fluctuations in the electric fields acting in the QD. The RTN process is modeled as in Ref.\cite{Testolin-2009,Mottonen-2006}, the noise fluctuates randomly between −1 and 1 with the frequency of the fluctuations controlled by the correlation time $1/\lambda$. Here, $\lambda$ is the frequency of jump times, where the jump time instants $\tau_{j}$ are
\begin{equation}
\tau_{j}=\sum_{k=1}^{j} -\frac{1}{\lambda} \ln(p_{k}),
\end{equation}
and the $p_{k}$ are random numbers with $p_{k} \in (0,1)$. The noise process $\eta(t)$ is described as
\begin{equation}
\eta(t)=(-1)^{\sum_{j}\Theta(t-\tau_{j})},
\end{equation}
where $\Theta$ is the Heaviside step function. 
For each step of the CNOT sequence, we consider the average effect of charge noise over the step sequence time $t_{i}$. We modeled the effect of the charge noise as a deviation of the exchange interaction value in the $i$-th step as 
\begin{equation}
\Delta J^{max}_{i}= \alpha \Big[ \frac{1}{t_{i}} \int_{\sum_{j=0}^{i-1} t_{j}}^{\sum_{j=0}^{i} t_{j}} \eta(t) dt \Big],
\end{equation}
where $\alpha$ is the coupling strength.
\begin{figure}[h]
	\begin{center}
		\includegraphics[width=0.8\textwidth]{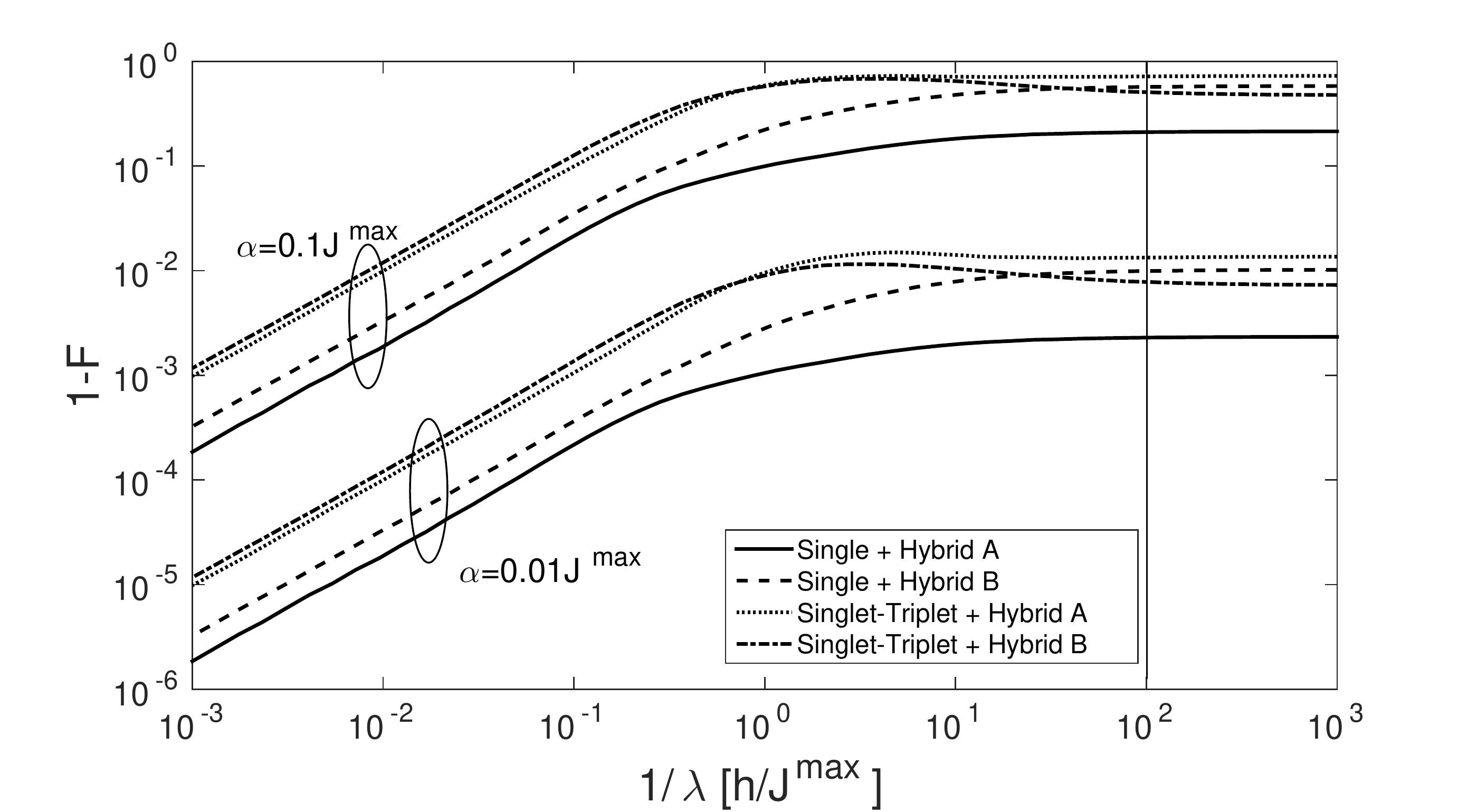}
	\end{center}
	\caption{\label{Fig:dJ} Gate infidelities for the CNOT gate sequences due to charge Noise as a function of the correlation time 1/$\lambda$ with $\alpha=0.01 J^{max}$ and $0.1J^{max}$. Vertical line marks a typical experimental $\lambda$ value, that when $J^{max}$=1 $\mu$eV gives $\lambda=0.01J^{max}/h\simeq$24.2 MHz.}
\end{figure}
Figure \ref{Fig:dJ} shows the gate infidelities $1-F$ of the CNOT gate sequences as a function of correlation time 1/$\lambda$ for the different systems considered with $\alpha=0.01J^{max}$ and $0.1J^{max}$. Systems have infidelity that generally increases when charge noise frequency $\lambda$ decreases. An apparent reduction in the infidelity is observed for singlet-triplet + hybrid B systems probably due to those particular CNOT sequences. Differently from Figs. \ref{Fig:TIE} and \ref{Fig:dEz}, configurations with single spin qubit exhibit lower infidelities in CNOT sequences than those with singlet-triplet qubit.   

\subsection{Combined effects}
Some considerations on the parameters of the noise model have to be done when an estimation of the combined effect of noises on the gate infidelity is required. The parameter values considered are: $J^{max}=$1 $\mu$eV as maximum exchange interaction value compatible with that derived in Ref.\cite{DeMichielis-2015}, $\sigma=0.25\times 10^{-3}h/J^{max}\simeq$1 ps as TIE standard deviation \cite{DatasheetPulseGenerator}, the hyperfine coupling value $\delta E_{z}$=3 neV (0.003$J^{max}$) for Si \cite{Assali-2011}, $\lambda=0.01J^{max}/h\simeq$24.2 MHz \cite{Mottonen-2006} as charge noise correlation frequency and $\alpha=0.01J^{max}=10^{-2} \mu$eV as coupling strength.

The resulting infidelities for the different systems are reported in Tab. \ref{Tab:1-Ffor4systems}.
\begin{table}[th]
	\caption{\label{Tab:1-Ffor4systems}
		Comparison of infidelities among the four systems with model parameters as reported in the text.}
	\begin{center}
		\begin{tabular}{c c | c}
			qubit L & qubit R & 1-F \\
			\hline
			Single & Hybrid A & 5.91 $\times$ $10^{-3}$ \\
			Single & Hybrid B & 1.55 $\times$ $10^{-2}$ \\
			Singlet-Triplet & Hybrid A & 1.34 $\times$ $10^{-2}$ \\
			Singlet-Triplet & Hybrid B & 7.86 $\times$ $10^{-3}$ \\
			\hline
		\end{tabular}
	\end{center}
\end{table} 

Under those conditions, system with a single spin qubit coupled to an hybrid qubit in configuration A exhibits the highest CNOT fidelity.   

\section{Conclusions}
Controlled-NOT gate sequences in mixed qubit architectures are presented. The double QDs hybrid qubit takes advantage of its only-exchange mechanism interaction to implement logical operations requiring only an electrical external control. This promising architecture is investigated when interconnection with different spin qubits, such as the QD single spin and the double QDs singlet-triplet qubits that assure longer coherence times, is considered. The search for the sequences for the two different interconnection schemes in the two different geometrical configurations A and B, due to the asymmetries in the two dots composing the hybrid qubit, was performed numerically developing and using a mixed simplex and genetic algorithm. The gate fidelity analysis is performed by taking into account time interval error, hyperfine coupling and charge noise sources. When TIE is considered, singlet-triplet with hybrid QD version B qubit shows lower CNOT infidelity with respect the other systems. Singlet-triplet paired to an hybrid QD version B qubit is again the best choice even when effects of hyperfine coupling induced magnetic noise are analyzed. Conversely, single coupled to hybrid qubit version A exhibits the lowest CNOT infidelity when charge noise arises. When all the considered noise sources are included altogether choosing real world parameters, our model shows that systems with a single spin qubit coupled to an hybrid qubit in configuration A exhibit the highest CNOT fidelity. 
The resulting noise effects in mixed qubit systems can be surely mitigated with more sophisticated control pulse arrangements such as dynamical decoupled gate sequences that will be presented in a future work. 

\begin{acknowledgements}
This project has received funding from the European Union's Horizon 2020 research and innovation programme under grant agreement No 688539.
\end{acknowledgements}

\appendix
\section{Effective exchange coupling constants}
In this Appendix, following the same procedure already exploited in Refs. \cite{Ferraro-2014,Ferraro-2015-qip}, all the detailed expressions for the exchange coupling constants between pair of electrons in both the mixed architectures considered are reported. The Schrieffer-Wolff effective Hamiltonian models (\ref{Hsingle}) and (\ref{Hst}) are derived by combining a Hubbard-like model with a projector operator method \cite{Schrieffer-1966}. As a result, the Hubbard-like Hamiltonian is transformed into an equivalent expression in terms of the exchange coupling interactions between pairs of electrons. Since the two dots composing the hybrid qubit are asymmetric, it follows that there are two different possible configurations for each architecture as shown in Figs. \ref{single_hybrid} and \ref{st_hybrid}. 

\subsection{Quantum dot single-spin qubit and double quantum dot hybrid qubit}
The expressions for the exchange coupling constants for the configuration A appearing in the effective Hamiltonian (\ref{Hsingle}) are given by
\begin{align}\label{couplingA}
&J_{1_R2_R}=\frac{1}{\Delta E_1}4(t_{1_R2_R}-J^{(1_R2_R)}_t)^2-2J^{(1_R2_R)}_e\nonumber\\
&J_{2_R3_R}=\frac{1}{\Delta E_2}4(t_{2_R3_R}-J^{(2_R3_R)}_t)^2-2J^{(2_R3_R)}_e\nonumber\\
&J_{1_R3_R}=\left(\frac{1}{\Delta E_3}+\frac{1}{\Delta E_4}\right)4J^{(1_R3_R)2}_t-2J^{(1_R3_R)}_e\nonumber\\
&J_{1_L1_R}=\frac{1}{\Delta E_5}4(t_{1_L1_R}-J^{(1_L1_R)}_t)^2-2J^{(1_L1_R)}_e\nonumber\\
&J_{1_L3_R}=\frac{1}{\Delta E_6}4(t_{1_L3_R}-J^{(1_L3_R)}_t)^2-2J^{(1_L3_R)}_e,
\end{align}
with the energy differences defined as
\begin{align}
&\Delta E_1=E_{(1,012)}-E_{(1,111)}\nonumber\\
&\Delta E_2=E_{(1,102)}-E_{(1,111)}\nonumber\\
&\Delta E_3=E_{(1,201)}-E_{(1,111)}\nonumber\\
&\Delta E_4=E_{(1,021)}-E_{(1,111)}\nonumber\\
&\Delta E_5=E_{(0,211)}-E_{(1,111)}\nonumber\\
&\Delta E_6=E_{(0,121)}-E_{(1,111)}\nonumber\\
\end{align}
where
\begin{align}
E_{(w,ijk)}=&w\varepsilon_{1_L}+i\varepsilon_{1_R}+j\varepsilon_{3_R}+k\varepsilon_{2_R}+ijU_{1_R3_R}+ikU_{1_R2_R}+\nonumber\\
&+kjU_{2_R3_R}+\delta_{i2}U_{1_R}+\delta_{j2}U_{3_R}+\delta_{k2}U_{2_R}+\nonumber\\
&+iwU_{1_L1_R}+jwU_{1_L3_R}.
\end{align}
The first index in parenthesis $w=0,1$ denotes the electron occupation for the single spin qubit $L$, while the indices $i,j,k=0,1,2$ denote the number of electrons in each level for the hybrid qubit $R$ ordered as depicted in Fig. \ref{single_hybrid}. The parameters involved are: the energy levels $\varepsilon_i$, the tunneling coefficients between different dots $t_{ij}$, the spin exchange $J_e^{ij}$ and the occupation-modulation hopping terms $J_t^{ij}$. 

Analogously the exchange coupling constants for the configuration B are defined as in Eq.(\ref{couplingA}) with the new inter-qubit interaction
\begin{equation}
J_{1_L2_R}=\frac{1}{\Delta E_5}4(t_{1_L2_R}-J^{(1_L2_R)}_t)^2-2J^{(1_L2_R)}_e
\end{equation}
where $\Delta E_5=E_{(0,112)}-E_{(1,111)}$. The energies corresponding to each configurations are given for the configuration B by
\begin{align}
E_{(w,ijk)}=&w\varepsilon_{1_L}+i\varepsilon_{1_R}+j\varepsilon_{3_R}+k\varepsilon_{2_R}+ijU_{1_R3_R}+ikU_{1_R2_R}+\nonumber\\
&+kjU_{2_R3_R}+\delta_{i2}U_{1_R}+\delta_{j2}U_{3_R}+\delta_{k2}U_{2_R}+\nonumber\\
&+kwU_{1_L2_R}.
\end{align}

\subsection{Double quantum dot singlet-triplet qubit and double quantum dot hybrid qubit}
The exchange coupling constants for the double QD singlet-triplet and the double QD hybrid qubits in configuration A appearing in the effective Hamiltonian (\ref{Hst}) are given by
\begin{align}\label{couplingAst}
&J_{1_R2_R}=\frac{1}{\Delta E_{1_R}}4(t_{1_R2_R}-J^{(1_R2_R)}_t)^2-2J^{(1_R2_R)}_e\nonumber\\
&J_{2_R3_R}=\frac{1}{\Delta E_{2_R}}4(t_{2_R3_R}-J^{(2_R3_R)}_t)^2-2J^{(2_R3_R)}_e\nonumber\\
&J_{1_R3_R}=\left(\frac{1}{\Delta E_{3_R}}+\frac{1}{\Delta E_{4_R}}\right)4J^{(1_R3_R)2}_t-2J^{(1_R3_R)}_e\nonumber\\
&J_{1_L2_L}=\left(\frac{1}{\Delta E_{3_L}}+\frac{1}{\Delta E_{4_L}}\right)4(t_{1_L2_L}-J^{(1_L2_L)}_t)^2-2J^{(1_L2_L)}_e\nonumber\\
&J_{2_L1_R}=\frac{1}{\Delta E_5}4(t_{2_L1_R}-J^{(2_L1_R)}_t)^2-2J^{(2_L1_R)}_e\nonumber\\
&J_{2_L3_R}=\frac{1}{\Delta E_6}4(t_{2_L3_R}-J^{(2_L3_R)}_t)^2-2J^{(2_L3_R)}_e,
\end{align}
where
\begin{align}
&\Delta E_{1_R}=E_{(11,012)}-E_{(11,111)}\nonumber\\
&\Delta E_{2_R}=E_{(11,102)}-E_{(11,111)}\nonumber\\
&\Delta E_{3_R}=E_{(11,201)}-E_{(11,111)}\nonumber\\
&\Delta E_{4_R}=E_{(11,021)}-E_{(11,111)}\nonumber\\
&\Delta E_{3_L}=E_{(02,111)}-E_{(11,111)}\nonumber\\
&\Delta E_{4_L}=E_{(20,111)}-E_{(11,111)}\nonumber\\
&\Delta E_5=E_{(10,211)}-E_{(11,111)}\nonumber\\
&\Delta E_6=E_{(10,121)}-E_{(11,111)}\nonumber\\
\end{align}
with
\begin{align}
E_{(wz,ijk)}=&w\varepsilon_{1_L}+z\varepsilon_{2_L}+wzU_{1_L2_L}+\delta_{w2}U_{1_L}+\delta_{z2}U_{2_L}+\nonumber\\
&+i\varepsilon_{1_R}+j\varepsilon_{3_R}+k\varepsilon_{2_R}+ijU_{1_R3_R}+ikU_{1_R2_R}+\nonumber\\
&+kjU_{2_R3_R}+\delta_{i2}U_{1_R}+\delta_{j2}U_{3_R}+\delta_{k2}U_{2_R}+\nonumber\\
&+izU_{2_L1_R}+jzU_{2_L3_R}.
\end{align}
The first (last) indices inside parenthesis, assuming only integer values between $0$ and $2$, denote the number of electrons in each level for qubit $L(R)$ ordered as depicted in Fig. \ref{st_hybrid}. 

For the configuration B the coupling constants are defined as in Eq.(\ref{couplingAst}) with the new inter-qubit interaction
\begin{equation}
J_{2_L2_R}=\frac{1}{\Delta E_5}4(t_{2_L2_R}-J^{(2_L2_R)}_t)^2-2J^{(2_L2_R)}_e
\end{equation}
where $\Delta E_5=E_{(10,121)}-E_{(11,111)}$. The energies are now given by
\begin{align}\label{a2}
E_{(wz,ijk)}=&w\varepsilon_{1_L}+z\varepsilon_{2_L}+wzU_{1_L2_L}+\delta_{w2}U_{1_L}+\delta_{z2}U_{2_L}+\nonumber\\
&+i\varepsilon_{1_R}+j\varepsilon_{3_R}+k\varepsilon_{2_R}+ijU_{1_R3_R}+ikU_{1_R2_R}+\nonumber\\
&+kjU_{2_R3_R}+\delta_{i2}U_{1_R}+\delta_{j2}U_{3_R}+\delta_{k2}U_{2_R}+\nonumber\\
&+kzU_{2_L2_R}.
\end{align}

\section{Graphical representation of CNOT gates}
In this Appendix a graphical representation of modulus and phase (gray scale) of the final transformation matrix for the CNOT gates for the four mixed architectures studied is shown. The resulting transformation matrices are obtained starting from the sequences reported in Tabs. \ref{Tab:SeqAndTimesCNOTv1A}, \ref{Tab:SeqAndTimesCNOTv1B}, \ref{Tab:SeqAndTimesCNOTv2A}, \ref{Tab:SeqAndTimesCNOTv2B}. The $4\times 4$ block in the up left corner corresponds to the CNOT matrix reported in Eq.(\ref{Eq:CNOT}).
\begin{figure}[h]
	\begin{center}
		\includegraphics[width=0.8\textwidth]{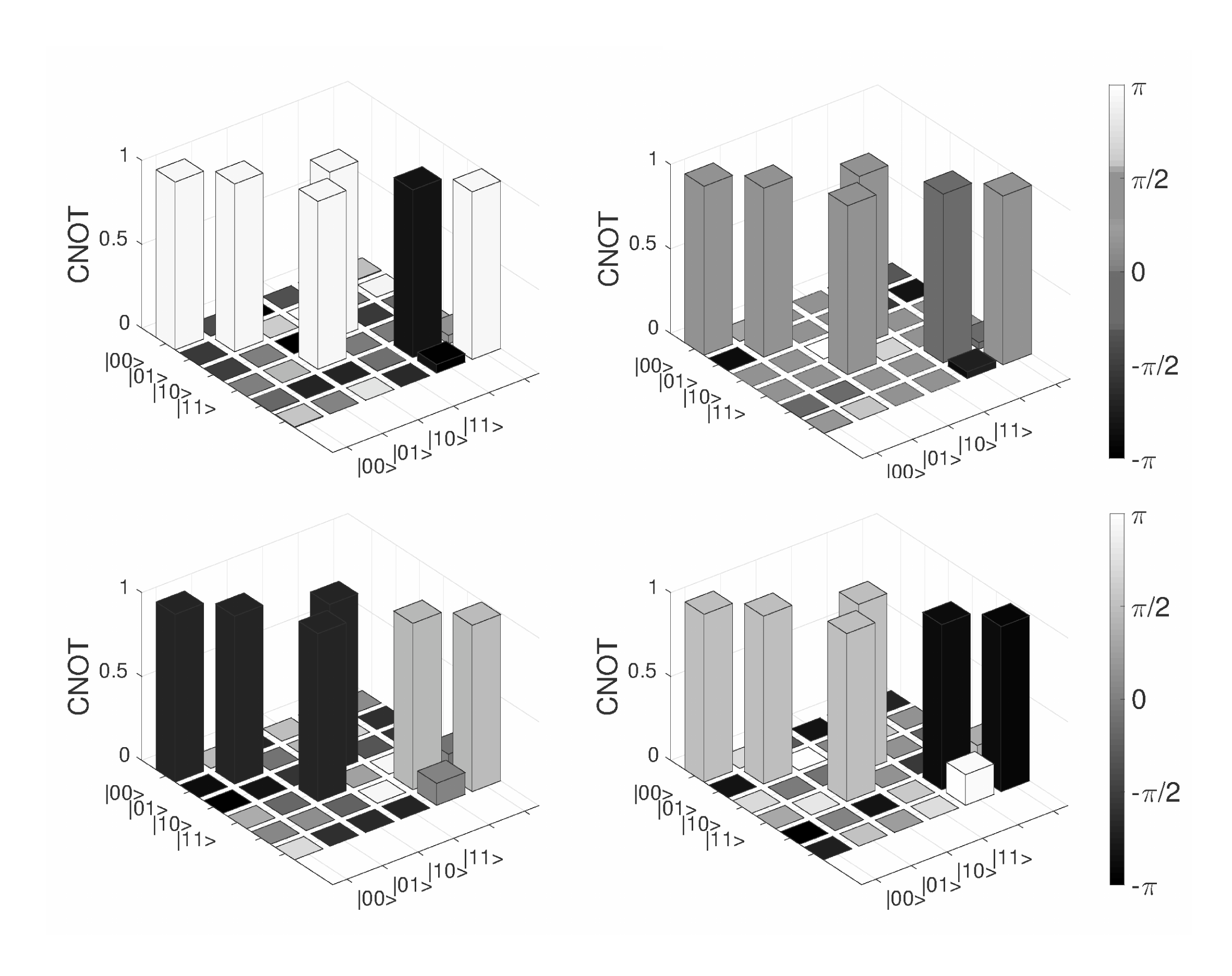}
	\end{center}
	\caption{\label{Fig:CNOT} Graphical representation of modulus and phase of the final transformation matrix for the CNOT gates. Top left (right): Single+Hybrid A (B); Bottom left (right): Singlet-Triplet+Hybrid A (B).}
\end{figure}


\bibliographystyle{spphys}       

\bibliography{Ref}

\end{document}